\documentstyle[aps,preprint]{revtex}

\newcounter{myfootnote}[page]
\newcommand{\email}[1]{
             \renewcommand{\thefootnote}{\alph{footnote}}%
             \setcounter{footnote}{\value{myfootnote}}%
             \footnote{#1}\stepcounter{myfootnote}
            }
\def\GeV{\,\mbox{\rm GeV}}
\newcommand{\be}{\begin{eqnarray}}
\newcommand{\ee}{\end{eqnarray}}
\newcommand{\ba}{{\bf p}_a}
\newcommand{\bc}{{\bf p}_c}
\newcommand{\bp}{{\bf p}}
\newcommand{\bP}{{\bf P}}
\newcommand{\bara}{{\bf \bar {\it a}}}
\newcommand{\barb}{{\bf \bar {\it b}}}
\newcommand{\barc}{{\bf \bar {\it c}}}
\newcommand{\bard}{{\bf \bar {\it d}}}
\newcommand{\barq}{{\bf \bar {\it q}}}
\newcommand{\barQ}{{\bf \bar {\it Q}}}
\newcommand{\nn}{\nonumber \\}
\def\j{$J/\psi \  $}
\def\phen{phenomenological }
\def\xf{$x_F$ }

\begin{document}
\tighten
\begin{titlepage}
\pagestyle{empty}
\vspace{1cm}
\title{\vspace{2cm}         \ \ \ \ \ \\
       \hspace{10cm} {\rm MPG-VT-UR 39/94 \\
       \hspace{10cm} GSI-94-75} \\
       \vspace{3cm}
        Quark exchange model for charmonium dissociation \\
        in hot hadronic matter}
\author{
K. Martins\email{email: martins@darss.mpg.uni-rostock.de}
and D. Blaschke\email{email:  blaschke@darss.mpg.uni-rostock.de}}
\address{Max--Planck--Gesellschaft AG "Theoretische Vielteilchenphysik"\\
Universit\"{a}t Rostock, D-18051 Rostock, Germany}
\author{E. Quack\email{email: quack@axp601.gsi.de}
}
\address{Theory Department, Gesellschaft f\"ur Schwerionenforschung (GSI), \\
Postfach 11 05 52, D-64220 Darmstadt, Germany
}
\maketitle

\begin{abstract}
A diagrammatic approach to quark exchange processes in
meson-meson scattering is applied to the case of inelastic reactions
of the type
$(Q\barQ)+(q\barq)\rightarrow (Q\barq) + (q\barQ)$, where $Q$ and $q$ refer
to heavy and light quarks, respectively.
This string-flip process is discussed as a microscopic mechanism
for charmonium dissociation (absorption) in hadronic matter.
The cross section for the reaction
$J/\psi + \pi \to D+ \bar D$ is calculated
using a potential model, which is
fitted to the meson mass spectrum.
The temperature dependence of the relaxation time for the \j
distribution in a homogeneous thermal pion gas is obtained.
The use of charmonium for the diagnostics of the state of hot hadronic matter
produced in ultrarelativistic nucleus-nucleus collisions is discussed.

\vspace{0.4cm}
\end{abstract}

\end{titlepage}
\newpage
%
\section{Introduction}

The interaction of a \j meson with strongly interacting matter
is to date still
a controversial subject. While the production of \j  can be
understood within perturbative QCD due to the large mass of the charm quark,
its further interaction with surrounding matter is essentially soft in
nature and as such not treatable perturbatively.
The knowledge of hadronic interactions, as well as their
modifications at finite temperature and density, is however
necessary for a proper
understanding of ultrarelativistic nucleus-nucleus collisions, especially
in view of a possible transition from hadronic to quark matter \cite{qm93}.
The suppression of \j was initially proposed as a signal for a quark-gluon
plasma \cite{matsui}. Such a suppression was observed by NA38 \cite{na38data}.
However, the data can be described by a variety of models
on a phenomenological basis, both in a plasma \cite{karsch}
and in a conventional hadronic scenario \cite{huefner,gavin2}.
Thus, the question of the significance of the \j signal remains as yet
undecided.

Plasma formation is not expected to occur in hadron--nucleus (hA) collisions.
However, data taken in hA collisions already show a considerable
reduction of \j production at low \xf (the region where \j is also measured
in nucleus--nucleus (AB) collisions) as compared to proton--proton ($pp$)
\cite{na3,alde}.
The suppression pattern in both hA and AB collisions
is found to be consistently described by a \phen absorption cross section of
$\sigma^{\psi N}_{\rm abs} \approx $ 5--7 mb \cite{gerschel1}.

On the other hand, it was recently argued that, due to the smallness of the
heavy quark--antiquark system ($Q \barQ$), a gluon needs to be sufficiently
hard in order to resolve this pair,
$Q_g^2 \ge 1/m^2_{\psi}$ \cite{kharzeev}. Only deconfined matter at
temperatures beyond the phase transition temperature $T_c$ contains
sufficiently many hard
gluons to cause a \j suppression of the observed magnitude. Matter in the
form of hadrons does not provide enough hard gluons, and consequently a
value of $\sigma^{\psi N}_{\rm abs} \approx $ 5--7 mb has been regarded as
unrealistic.
This obvious contradiction is one example for the need of an understanding
of hadron--hadron
interactions on a more fundamental level.

With the present work, we aim to
provide a step towards filling this gap.
The approach we use is the description of hadrons as bound states
of quarks. It allows one to consistently account for substructure effects.
A full treatment of the hadron--hadron interaction is to date
not possible due to the non--perturbative character of QCD in this region.
A microscopic calculation of the \j  breakup process by impact ionization
has been performed within perturbative QCD in Ref. \cite{wittmann} for a dense
partonic environment and in Ref. \cite{kharzeev} for a hadronic medium.
However, in these approaches non--perturbative correlations in the
final state (charmed hadrons) have been neglected, i.e.~only the breakup of
a \j into free charm quarks was considered. As lattice gauge simulations
of QCD suggest
\cite{lattice1,lattice2,koch-brown}, hadronic correlations persist even for
temperatures well above the deconfinement transition.
Therefore, effective approaches to \j dissociation in the non--perturbative
domain of strongly correlated quark matter consider
this process as a quark exchange (string--flip) process \cite{blaschkeroepke}.
The role of quark exchange processes in hadron-hadron interactions has
been investigated in  several approaches \cite{brod,qex1,qex2,qex3,qex4}.
Recently, a systematic analysis of quark exchange
contributions to the meson-meson interaction has been given
in Refs. \cite{barnes,quarkexchange} within a diagrammatic technique.
These approaches use a non-relativistic quark potential model to describe
mesons as bound states. They have been applied to the elastic scattering of
light mesons.
When translating the diagrams into the language of Green
functions \cite{quarkexchange}, a generalization to finite temperatures
and densities  as well as to a relativistic effective meson
theory is possible.

In the present work, we extend this diagrammatic technique to the
calculation of the cross section for the inelastic reaction
$(Q\barQ)+(q\barq)\rightarrow (Q\barq) + (q\barQ)$, where $Q$ and $q$ stand
for heavy and light quarks, respectively.
We consider the process of charmonium dissociation by inelastic collisions
with light mesons. In particular, we calculate the cross section of the
reaction  \j + $\pi \longrightarrow D^* + \bar{D}$ as a function of the
relative kinetic energy of the mesons and address its application to the
analysis of the kinetics of charmonium dissociation in heavy ion collisions.

The section \ref{ex} gives the general formalism for inelastic meson--meson
scattering. In section \ref{diss}, the special case of charmonium
dissociation is
treated within this formalism, and cross sections for the main processes
are calculated.
These are then used in section \ref{kinetics} to study the absorption in
a pion gas.
In section \ref{experiment} we discuss the  situation encountered
in the experiment.
\section{Quark exchange contribution to meson-meson scattering}
\label{ex}
In this section, we present the formalism for the calculation of the cross
section of quark exchange processes between mesons. We mainly follow the
notation of Ref.\@ \cite{barnes}.
We consider the two--meson scattering
process
\nopagebreak{
$A (a\bar{a}$) + $B (b\bar{b})
\to C (a\bar{b}) + D (b\bar{a})$,
}
where the interchange of the quark content (in brackets) corresponds to
a flavor rearrangement.
This  process
dominates the cross section behavior at low relative energies of the mesons,
while at higher energies the additional production of light $q\bar{q}$ pairs
sets in which is not contained in the present approach.

The differential cross section for the process $i \to f$ is given by
\be
   \frac{d\, \sigma_{fi}(s,t)}{d\, t}
   &=& \frac{1}{64 \pi s} \frac{1}{{ P}^2(s)} |{\cal M}_{fi}(s,t)|^2,
\label{sigma_allgemein}
\ee
where $P(s)$ is the
relative three--momentum of incoming particles in their center of mass frame.
For the relation between $P$ and $s,t$ see App.\@ \ref{appendixa}.
The indices $i$ and $f$ stand for the initial and final two--meson
states.
The central problem is the calculation of the relativistic invariant
matrix element ${\cal M}_{fi}(s,t)$.
For this, effective theories have to be used in the low energy domain, where
perturbative QCD is not applicable.
One specific property of hadron--hadron scattering is the
color neutrality of asymptotic states, such that a single one gluon exchange
between hadrons is forbidden.
The quark exchange process, however, is possible and the matrix element
reads in Born approximation \cite{barnes,quarkexchange}
\be
  {\cal M}_{fi} &=&
 {\cal N}  \left<\Psi^A \Psi^B\right|{H}_{AB,CD}\left|\Psi^C \Psi^D \right>,
\label{Mfi}
\ee
with the meson--meson interaction Hamiltonian $H_{AB,CD}$ and
a product ansatz for the incoming  (outgoing) two--meson states formed
by the mesons $A,B$ ($C,D$).
Four--momentum conservation is implemented in this matrix element.
The normalization factor ${\cal N}$ is needed in order to get the correct
form of ${\cal M}_{fi}$ from the nonrelativistic transition matrix element.
With our convention, it reads
\be
{\cal N} = \frac{1}{\Omega_0}
\prod_{i=A\ldots D} \sqrt{2E_i \Omega_0}\, ,
\ee
where $\Omega_0$ is the normalization volume of the states $\Psi$,
which is set in the following to unity, and $E_i=\sqrt{m_i^2+p_i^2}$.
The calculation of ${\cal M}_{fi}$ is performed in the center of mass frame of
the mesons A and B. The resulting differential cross section  $d\sigma/d\, t$
is expressed in
terms of Mandelstam variables, that is, in Lorentz invariant form.
The total cross section for scattering into channel $f$ is obtained
by integrating over $t$
\be
\sigma_{fi}(s)
   &=&\int_{t_-}^{t_+} dt \frac{d\, \sigma_{fi}(s,t)}{d\, t} \; ,
\label{sigma(s)}
\ee
where $t_+ (t_-)$ is the maximal (minimal) possible momentum transfer $t$.
The  $t$ integration can be transformed into an integration over
$z=\cos \vartheta (\bP,\bP')$, where $\vartheta (\bP,\bP')$ is
the  angle between the relative momenta $\bP$ and $\bP'$ of incoming and
outgoing mesons, respectively.
For nonidentical particles the following relation holds:
\be
\label{sigmaifs}
 \sigma_{fi}(s)
       &=&\frac{1}{32 \pi s}\frac{P'(s)}{P(s)}
           \int_{-1}^1 d z\,|{\cal M}_{fi}(P(s),P'(s),z)|^2.
\ee
\subsection{The quark exchange Hamiltonian}
It has been shown in Refs.\@ \cite{barnes,quarkexchange}, that in the quark
potential model the
Hamiltonian of the quark exchange process in meson--meson scattering
can be represented as an effective two--quark interaction
followed by a quark interchange between the mesons.
The result of the diagrammatic analysis of all topological inequivalent
contributions to the quark exchange matrix element is shown  in Fig.\@
\ref{feyn3}.
According to Eq.\@ (\ref{Mfi}) the matrix element of the
quark exchange Hamiltonian in the four quark basis reads
\be
 \left<a,\bara,b,\barb|  H_{AB,CD}| c, \barc,d,\bard \right> =&
     \sum\limits_{{i=a, \bar a}\atop {j=b, \bar b}}
& \left<a,\bara,b,\barb|  H^I_{ij}| c, \barc,d,\bard \right>.
\label{H-I}
\ee
For illustration we give the first of these four terms
\be
 \left<a,\bara,b,\barb| H^I_{a \barb}| c, \barc,d,\bard \right>
    &=&
   \sum_{{a',\bara ' }\atop {b',\barb '}}
  \left<a,\barb|H^I|a',\barb ' \right>\left<\bara, b|\bara ',b' \right>
  \left<a', \bara ',b', \barb '|c,\barc,d,\bard\right>,
\ee
where $a\ldots \bard' $ denote three--momentum, spin, flavor and
color quantum numbers of the quark or antiquark
($a=\{\bp_a,{\bf s}_a, f_a, c_a\}$).
The last bracket selects those contributions in the sum over all quark quantum
numbers, which match to the final state two--meson wave function.
The other terms in Eq.\@ (\ref{H-I}) are obtained in an analogous manner,
where the interaction acts between the particles  $i$ and $j$.

In the quark potential model, the two--quark interaction of
a meson is
given by the interaction Hamiltonian $H^I$ of Fermi-Breit type.
The same interaction is assumed to act also between the quarks of
different mesons.
It consists of the usual kinetic term, a nonrelativistic potential
($H^V$) and relativistic  corrections arising from
spin-spin ($H^{SS}$) and
spin-orbital ($H^{LS}$) interaction, a tensor interaction ($H^T$) and a
spin independent term ($H^{SI}$), see \cite{lucha} for a review.
In the present work only S wave mesons are considered.
Thus the two--quark interaction Hamiltonian $H^I$ is a sum of the
quark--quark potential $H^V$ and the spin--spin interaction $H^{SS}$ only,

\be
   \left<i,j|H^I|i',j'\right>
   &=&   \left<i,j|H^V|i',j'\right>+   \left<i,j|H^{SS}|i',j'\right>.
\label{breitfermi}
\ee
For the reason of mathematical tractability of the matrix element (\ref{Mfi})
we choose an effective Gaussian ansatz for the orbital part of interaction,
which in momentum space reads
\begin{equation}
\left<a, \barb|H^V|a',\barb '\right>=-V_0(8 \pi x)^{3/2}
       \text{e}^{-2x(\bp_a-\bp'_{a})^2}
      \delta_{a,a'}^{_{(S,F,C)}} \delta_{\barb,\barb'}^{_{(S,F,C)}}
      \delta_{\bp_a+\bp_{\bar b},\bp'_a+\bp'_{\bar b}},
\label{ImpPot}
\end{equation}
with parameters $V_0$ and $x$ for the depth and range.
 A Kronecker symbol with superscript $S,F$
or $C$ is understood to act in spin, flavor or color spaces, respectively,
e.g.~
$\delta_{j,j'}^{_{(F,C)}}=\delta_{{\bf s}_j,{\bf s}_{j'}}\delta_{f_j,f_{j'}}
    \delta_{c_j,c_{j'}}$.
This potential does not account for confinement.
The use of a nonconfining
potential follows Ref.\@ \cite{barnes} and is justified as long as the wave
functions
become small in the vicinity of the edge of the potential. In particular for
the mesons containing heavy quarks, this condition holds.
The spin--spin interaction is taken in the standard form  \cite{lucha}
\be
  \left<i,j|H^{SS}|i',j'\right>&=&
        \frac{32 \pi \alpha_s}{9 m_i m_{j}}
       {\bf s}_i {\bf s}_{j} \delta_{i,i'}^{_{(F,C)}}\delta_{j,j'}^{_{(F,C)}}
      \delta_{\bp_i+\bp_{j},\bp'_i+\bp'_{j}},
\label{HSS}
\ee
where $m_i$ and $m_j$ are the constituent quark masses.
This can be rewritten in terms of $H^V$ (Eq.\@ \ref{ImpPot}) since the
identity operator in momentum space can be understood as a limit of
the Gaussian potential  for $x\to 0$ and
$V_0\to (8 \pi x)^{-3/2}$.
%
\subsection{The meson wave functions}
We decompose the mesonic wave functions into orbital ($\Phi$),
spin ($\chi_S$), flavor and color ($\chi_{FC}$) parts,
  \be
\label{Psiansatz}
   \big|\Psi^A_{\bP_A}\big>&=&|\Phi^A_{\bP_A}\big>\otimes |\chi_{S}^A\big>
               \otimes |\chi_{FC}^A\big>,
  \nn
   \big<a,\bara  \big|\Psi^A\big>
 &=&\Phi^A_{\bP_A}(\ba,\bp_{\bara }) \chi_{S}^A({\bf s}_a,{\bf s}_{\bara})
 \chi_{FC}^A(f_a, f_\bara,c_a,c_\bara).
\ee
Instead of finding the exact eigenfunction of the two particle Schr\"odinger
equation we use trial Gaussian wave functions and find the best
approximation by using the Ritz variational principle.
The orbital part of $1S$ state wave function is given by
\be
\label{1S}
    \Phi^A_{\bP_A}({\bp_a},{\bp}_{\bara })&=&(2\pi)^{3/2}
      \left(\frac{4 \lambda_A}{ \pi}\right)^{3/4}
         \exp{\left[-2 \lambda_A \tilde \bp_A^2\right]}
           \delta_{\bP_A,(\bp_a+\bp_{\bara })},
 \ee
where $A$ stands for the quantum numbers and
${\bf P}_A=\bp_a+\bp_{\bara }$ for the total momentum of meson $A$.
The relative momentum of the quark and antiquark in
the meson is $\tilde \bp_A=\eta_A \ba -(1-\eta_A) \bp_{\bara }$, where
      $\eta_A={m_{\bara }}/(m_a+m_{\bara })$.
The constant $\lambda_A$ is related to the mean squared meson radius via
$\left<r^2\right>_A= 6 \lambda_A$.

We calculate the matrix element in the center of mass frame of mesons
A and B, where $\bP_A+\bP_B=\bP_C+\bP_D=0$ because of total momentum
conservation. Let us introduce the notation
\be
         \bP=&\bP_A=&-\bP_B,
  \nn
         \bP'=&\bP_C=&-\bP_D.
\label{comdef}
\ee
The generalization of the wave functions to excited states is straightforward.
If one would also consider P waves, the spin-orbit and tensor terms of the
interaction Hamiltonian had to be taken into account additionally.
The parameters of potential and wave functions are fitted to the
masses of the $\pi , \rho , J/\psi , \psi' , D$ and $D^*$ mesons,
see App. \ref{appendixb}.
\subsection{The transition matrix element}
According to the diagrammatic analysis of the contributions to the
matrix element (\ref{Mfi}), there are four
contributions to be evaluated, see Fig.\@ \ref{feyn3}.
The first two diagrams  correspond to the so called
capture diagrams of Ref.\@ \cite{barnes}, since the interacting quarks
are captured in one meson in the final state.
 The others represent the transfer diagrams.
The additional diagrams that arise,
if identical quarks are present in the considered process,
have the same amplitude and thus can be accounted for by a factor 2.

Since $H^I$ is a sum of an orbital and an spin--spin interaction, the
transition matrix element can be written as a superposition
\be
  {\cal M}_{fi}&=& \sum_{{i=a, \bar a} \atop {j=b, \bar b}} {\cal M}^V_{ij}
                     + {\cal M}^{SS}_{ij}.
\label{Mfisumme}
\ee
We make use of the product ansatz for the wave functions
Eq.\@ (\ref{Psiansatz})
and calculate the contributions from the terms of  Eq.\@ (\ref{Mfisumme}) to
the matrix element ${\cal M}_{fi}$.
Each of these  factorizes into an
orbital ($I_O$), spin ($I_S$)
and flavor--color ($I_{FC}$) part.
$H^V$ acts on the orbital part of the wave functions, and $H^{SS}$ on
the spin part.
For the matrix element ${\cal M}^V_{ij}$ we obtain
\be
\label{Va-b}
   {\cal M}_{ij}^{V}(P,P',z)
      &=&{\cal N}(P,P') I^V_{O,ij}(P, P',z)\,
            I_{S,ij}^{V}\, I_{FC,ij}^{V},
\ee
with
\be
 I^V_{O,ij}(P,P',z)&=&\big<\Phi^A_\bP \Phi^B_{\text{-}\bP}\big|
                     H^V_{ij}\big|\Phi^C_{\bP'} \Phi^D_{\text{-}\bP'} \big>,
\\
  I^V_{S,ij}&=&\big<\chi^A_S \chi^B_S\big|\chi^C_S \chi^D_S \big>,
\\
  I^{V}_{FC,ij}&=&      \big<\chi^A_{FC} \chi^B_{FC}\big|
                       \chi^C_{FC} \chi^D_{FC} \big>.
\ee

The calculation of the orbital, spin  and flavor-color factors is explained
in App.\@ \ref{appendixc}.
Here we only give the result:
\be
{\cal M}^{V}_{a\barb}(P,P',z)
   &=&-{\cal N}(P,P') \,I_{S,a \barb}^V I_{FC, a \barb}^V K_{a \barb}
    {\exp}{\left[-\left(\alpha_1 P^2+\alpha_2 {P'}^2
                +\alpha_3 P'Pz\right)\right]},
\label{MfiLambdaCapt}
\\
{\cal M}^{V}_{a b}(P,P',z)
   &=&-{\cal N}(P,P') \,I_{S,a b}^V I_{FC, a b}^V K_{a b}
    {\exp}{\left[-\left(\beta_1 P^2+\beta_2 {P'}^2
                +\beta_3 P'Pz\right)\right]}.
\label{MfiLambdaTrans}
\ee
The constants $\alpha_1, \cdots, \beta_3, K_{a \barb}$
and $K_{ab}$ are fixed by the parameters of the potential  ($V_0, x$) and
 the wave function ($\lambda_i$)  and
are explained in Eqs.\@ (\ref{alphas})--(\ref{lambdabeta}).
It can be shown that the ${\cal M}_{\bara  b}$ diagram is obtained from
${\cal M}_{a \barb}$ by
interchanging
mesons $C$ and $D$ and replacing $z$ by $-z$ in
(\ref{ortsantappendixabarb}).
Thus, one can express the matrix element ${\cal M}^V_{\bara  b}$ in terms
of ${\cal M}^{V}_{a\barb}$ by exchanging
$\eta_C \leftrightarrow 1-\eta_D$ and $\lambda_C \leftrightarrow \lambda_D$
and inserting the corresponding spin and flavor--color factors $I^V_S$ and
$I^V_{FC}$.
The same relations are valid between
 ${\cal M}^V_{a b}$ and
 ${\cal M}^V_{\bara \barb}$ .
$P'$ is fixed by $P$ due to energy conservation.

The corresponding matrix elements from the spin--spin interaction are
\be
   {\cal M}_{ij}^{SS}(P,P',z)
      &=&{\cal N}(P,P') I^{SS}_{O,ij}(P,P',z)\,
            I_{S,ij}^{SS}\, I_{FC,ij}^{SS}\, ,
\ee
with
\be
 I^{SS}_{O,ij}(P,P',z)&=&\big<\Phi^A_\bP \Phi^B_{\text{-}\bP}\big|
                     \Phi^C_{\bP'} \Phi^D_{\text{-}\bP'} \big>,
\\
  I^{SS}_{S,ij}&=&\big<\chi^A_S \chi^B_S\big|H^{SS}_{ij}\big| \chi^C_S \chi^D_S
\big>,
\\
I^{SS}_{FC,ij}&=&I^V_{FC,ij}.
\ee
In this case, the Hamiltonian acts on the spin part of the wave functions.
The spin factors $I^V_{S}$ and $I^{SS}_S$ are given in Table \ref{ispintable}.
Because of the $1/(m_a m_{\bar b})$ dependence, the spin--spin interaction
$H^{SS}$ dominates the matrix element when the interacton of light mesons
is considered.
In Ref.\@ \cite{barnes}, the orbital interaction $H^V$ has been
disregarded in the calculation of the $\pi^+ \pi^+$ scattering phase shifts,
see also the following subsection.
For our present application to the charmonium dissociation process,
the contribution of $H^V$ plays the dominant role in the transition matrix
element (\ref{Mfisumme}) and it will be examined in more detail in section
 \ref{diss}.
\subsection{Elastic $\pi^+ \pi^+$ scattering}
In this paragraph, we give the instructive limiting case of
four equal quark masses $m_a=\ldots =m_{\bard}=m_q$. That is, we consider the
scattering of identical spinless 1S mesons with masses $m$, described by
$\lambda_A=\ldots =\lambda_D=\lambda$. In this case the absolute values
of the incoming and outgoing relative momenta $P$ and $P'$ are equal.
We multiply by a factor of 2 in order
to take into account the diagrams which arise in addition to those containing
distinguishable particles.
In this case, the matrix elements
get a  transparent form. The spin-spin term reads
\be
   {\cal M}^{SS}(P,z)&=& \frac{32 \pi \alpha_s}{9 m_q^2}
                s  \Bigl\{
                -2\left(\frac{4}{3}\right)^{3/2}
                 \exp{\left[-4 \lambda /3 P^2\right]}\Bigr.
  \nn
               & &+ \Bigl. \exp\left[-\lambda P^2(1+z)\right]
                +\exp\left[-\lambda P^2(1-z)\right]\Bigr\}.
\ee
It has been shown in \cite{blaschkeroepke,barnes}, that the low energy
scattering phase shifts of $\pi^+\pi^+$ scattering can be well described
by this matrix element.
The  minimal relativistic approach to quark exchange processes
in hadron--hadron scattering has also proven successful in the description
of $K\pi$ and $KN$ scattering \cite{barnes2}. We expect that in processes,
where quark creation and annihilation is negligible, the presented
approach will be applicable.
\section{Charmonium dissociation}
\label{diss}
In this section, we apply the presented formalism to the specific
case of the breakup of \j when scattering on hadrons. We calculate the
absorption cross section from the quark exchange process, which is a function
of the relative kinetic energy of the two scattering mesons,  and discuss
the implications for realistic physical situations.
To demonstrate the importance of correlations in initial and final states,
we also discuss the breakup reaction of \j into free quarks. Here, the result
corresponds to previous perturbative calculations \cite{kharzeev}.

Charmonium absorption processes in hadronic matter have been considered
in several works, e.g.\@ \cite{gerschel1,vogtsystematics,prorok3}.
Basic processes for charmonium dissociation in hadronic matter are
\be
a)&\  J/\psi+\pi \to D(1S)+ \bar D(1S)  \quad
&\Delta m \ge 0.643 \GeV,
  \nn
b)&\  J/\psi+\rho \to D(1S)+ \bar D(1S) \quad
&\Delta m \ge -0.13 \GeV,
  \nn
c)& \  J/\psi + N \to \Lambda_c+\bar D(1S) \hspace{2.5em}
&\Delta m       \ge 0.258 \GeV.
\label{abc}
\ee
Generically, we denote
$D^+,D^-$ or $D^0$ as $D$, and correspondingly $\bar{D}$ for the antiparticles.
$D(1S)$ represents either $D$ or $D^*$.
Note that the reaction $J/\psi+\pi\to D+ \bar D$, without excited final
states, is forbidden by angular momentum conservation. The reaction thresholds
for the possible processes are given by the mass differences $\Delta m$.
All these reactions are examples of inelastic quark exchange processes among
hadrons. In the following, we work out our formalism considering process
$a)$ which describes charmonium absorption in a pion gas. Other processes
including higher meson states such as $\chi_c$ and $\psi'$ can also be
considered, see the discussion in the conclusions.
Process $c)$ describes \j absorption on nucleons and can be treated on
a similar basis.
\subsection{\j absorption by pion impact}
We apply the approach given in section \ref{ex}
to calculate the energy dependent cross section of the process a) by
specifying the initial mesonic states $A = J/\psi \, (Q\barQ),\,
B = \pi \,(q \barq )$ and the final state ,
$C = D(1S)\, (q \barQ), \, D = \bar D(1S)\, (Q \barq )$, where $Q$ is the
heavy charm quark and $q$ the light u or d quark.

In order to work out the details in a transparent way, we use Gaussian wave
functions and a Gaussian shape for the interaction potential which binds the
quark-antiquark pairs into mesons.
With the parameters of App.\@ \ref{appendixb}, we obtain a satisfying
description of the relevant meson spectrum, see Table \ref{masstable}.
The choice of the Gaussian class of functions
 has the advantage that the calculation of the cross
sections can be performed analytically, which makes the results more
transparent.
Eq.\@ (\ref{MfiLambdaCapt}) and (\ref{MfiLambdaTrans}) are now used to
calculate the cross section for the charmonium dissociation reaction
$J/\psi+ \pi \to D(1S)+ \bar{D}(1S)$. Due to the large charm mass,
the spin-spin interaction is negligible and we keep only
the potential interaction.
Then, for each final state channel, four matrix elements
have to be computed. The integral over $z$ in Eq.\@ (\ref{sigmaifs})
can be performed analytically with the result
\be
  \int_{-1}^1 d \,z\,|{\cal M}_{fi}(P,P',z)|^2
       &=&|I^V_{S}\, I^V_{FC}|^2 4
          {\cal N}^2(P,P')
      \nn
         &\times& \left\{
            K_{a \bar b}^2 \exp\left[-2\alpha_1P^2-2\alpha_2 P'^2\right]
  \left(1+\frac{\sinh(2\alpha_3 P' P)}{2 \alpha_3 P' P}\right) \right.
      \nn
          &&
             +K_{a b}^2 \exp\left[-2\beta_1P^2-2\beta_2 P'^2\right]
  \left(1+\frac{\sinh(2\beta_3 P' P)}{2 \beta_3 P' P}\right)
      \nn
          &&
     -2 K_{a \barb} K_{ab}
          \exp\left[-(\alpha_1+\beta_1)P^2-(\alpha_2+\beta_2){P'}^2\right]
          \nn
       &&\times \left. \left(\frac{\sinh (\alpha_3+\beta_3)P'
P}{(\alpha_3+\beta_3)P'P}
             +\frac{\sinh (\alpha_3-\beta_3)P' P}{(\alpha_3-\beta_3)P'P}
          \right)\right\}.
\label{zInt}
\ee
Here, the first term arises from the two capture diagrams. They have the same
spin and flavor-color factor and differ only in the sign of $z$, thus
${\cal M}_{a \barb}(z) = {\cal M}_{\bara  b}(-z)$ and consequently
$K_{a \barb} = K_{\bara  b}$. In an analogous manner, we have
${\cal M}_{a b}(z) = {\cal M}_{\bara \barb}(-z)$ and
$K_{a b} = K_{\bara  \barb}$ for the transfer diagram that corresponds to
the second term of the equation above.
The last term contains an interference of both processes.
We have defined $I^V_S:=I^V_{S,a \barb}=I^V_{S,\bara  b}=I^V_{S,a b}=
I^V_{S,\bara \barb}$ and $I^V_{FC}:= I^V_{FC,a \barb}=I^V_{FC,\bara  b}=
-I^V_{FC,a b}=-I^V_{FC,\bara \barb}$.
{}From the parameters of the potential model (\ref{parameter}), we obtain the
values $\alpha_1$=$1.37 \GeV^{-2}, \alpha_2$=$1.22GeV^{-2},
\alpha_3$=$0.059 \GeV^{-2}$, $\beta_1$=$0.717\GeV^{-2},$
$\beta_2$=$0.507\GeV^{-2}$ and $\beta_3$=$0.368\GeV^{-2}$.
Inserting this result in (\ref{sigmaifs}), with ${P}(s)$ and ${P'}(s)$
from Eq.\@ (\ref{pcom}),
 we obtain the cross section $\sigma_{fi}$ for a specific
final state $f$. For the total \j breakup cross section due to pion impact,
we sum the  possible final state combinations of low lying D mesons to get
\be
       \sigma_{\text{abs}}(s)&=& \sum_{f=1}^4\sigma_{fi}(s),
\ee
where $s$ is the center of mass energy of the \j and $\pi$.
The resulting \j absorption cross section,
which is a function of the relative kinetic
energy of \j and $\pi$ in the c.m. system, $E^{cms}_{rel}=s-(m_\psi+m_\pi)^2$,
is the central
result of this section. We show it as a
function of $E^{cms}_{rel}$ in Fig.\@ \ref{sigmabild}. Here, the parameter
values from Eq.\@ (\ref{parameter}) are used and all low--threshold processes
according to Table \ref{channels} are included except the lowest D\={D}
channel which is forbidden by angular momentum conservation.

The behavior of the cross section is characterized by a threshold at
$s_0=(m_C+m_D)^2$ and a strong enhancement near
this threshold, as well as
an exponential fall-off towards higher energies. We obtain a peak value of
about 15 mb at  $E^{cms}_{rel}=1 \GeV$.
The absorption cross section is approximately described
by the fit formula
\be
\label{fit}
   \sigma^{\text{fit}} (s)&\cong &\sigma_0 \cdot \left(1-\frac{s_0}{s}\right)^2
                \exp \left[-a(s-s_1)\right]\theta(s-s_0).
\ee
The fit parameters for different possible final states
$\sigma_0, a$ and $s_1$ are given in Table \ref{channels}.

As we mentioned in the beginning, we do not consider the
inelastic production of additional light $q\bar{q}$ pairs, which sets in at a
threshold of $\sqrt{s_0}+m_q+m_{\bar{q}}$.
Therefore, the exponential decrease in Eq.\@ (\ref{fit}) is not considered to
be realistic in view of the additional final state channels opened beyond this
energy.

\subsection{Phenomenology of hadron--hadron cross sections}
The large value of the absorption cross section we obtained within our
calculation is at first sight a rather unexpected result.
However, what was calculated is the cross section
of a preformed, full--size \j on a $\pi$. This is in most situations not
realistic. In real life, the $Q \barQ$ pair expands from a
small object at the creation vertex to its full size \cite{frankfurt,blahuef}.
 The initial size can be estimated to be
$\left< r^2 \right>^{1/2}_{Q\barQ} \sim 1/(2m_c) \sim 0.06$ fm,
as supported by charmonium hadroproduction and photoproduction experiments.
 A quantum mechanical treatment of the expanding $Q\barQ$
state which simultaneously interacts with hadrons gives a time scale of this
expansion of $\tau_{exp}^{Q\barQ}$ = 0.85 fm in the $Q\barQ$ rest frame
\cite{quack}.

In the present hA experiments, the kinematics are such that asymptotic
$J/\psi$'s are observed only at high momenta in the final state. Then,
$\tau_{exp}^{Q\barQ}$ has to be multiplied by a rather large $\gamma$ factor.
In other words, the \j is only formed far outside the nucleus.
This has two consequences.
Firstly, the $Q\bar{Q}$ interacts inside the nucleus still as a correlated,
but considerably small state.
We investigate this situation by describing the
initial $Q\bar{Q}$ state with a wave function narrower than the one of the
$J/\psi$, which is done by changing the wave function parameter
$\lambda_{Q \barQ}$ accordingly.
What we find is a decrease of the breakup cross section with decreasing
$Q\bar{Q}$ size. More precisely,
\be
\label{povhhuefner}
 \sigma_{\text{abs}} \propto \left< r^2 \right>_{Q\barQ} \; .
 \ee
This confirms the the phenomenological Povh--H\"ufner relation \cite{povh}
of hadron--hadron cross sections.
Therefore, in realistic experimental situations, the cross section of 15 mb is
lowered according to the kinematical circumstances.
Secondly, possible differences in the final state interaction of \j and
$\psi'$,
as expected already from their difference in size, are delayed, and thus
become invisible because the difference in their asymptotic states appears
only after they have
left the target nucleus. This is supported by the experimental observation
of an identical depletion of \j and $\psi'$ in heavy nuclei \cite{alde}.
\subsection{\j breakup without final state correlations}
As mentioned in the introduction, it was argued recently within a perturbative
approach that a \j breakup reaction via gluon exchange requires a relatively
hard gluon in order to resolve the small $Q\bar{Q}$ state \cite{kharzeev}.
The result we obtain for the cross section, Fig.\@ \ref{sigmabild},
is completely
different from the cross section obtained in such a perturbative approach.
However, the quark exchange process we considered is also very different from
a gluon exchange and essentially nonperturbative in nature.
We note in this context that the present treatment can be traced back to older
works of Gunion,
Brodsky and Blankenbecler on composite models of hadrons
\cite{brod}. They showed that even in certain short--range interactions
constituent exchange dominates over gluon exchange processes.

In order to illustrate this important point in the context of our approach,
we calculate the cross section for the breakup reaction of \j and  $\pi$
into four asymptotically free quarks within our effective model. Instead of
the Gaussian wave functions (\ref{1S}) we define the final state as plane
waves, in momentum space representation
\be
\Phi^C_{\bP'}(\bp_c,\bp_\barc)&=&\delta_{\bP',(\bp_c+\bp_{\barc})}
                          \delta_{(\bp_c-\bp_{\bar c})/2,\tilde\bp_C},
\nn
\Phi^D_{\text{-}\bP'}(\bp_d,\bp_\bard)&=&\delta_{\text{-}\bP',
                         (\bp_d+\bp_{\bar d})}
                          \delta_{(\bp_d-\bp_{\bar d})/2,\tilde \bp_D}.
\ee
The final four quark state is defined by the momenta $\tilde\bp_C,\tilde\bp_D$
and $\bP'$. The state $C$ contains the charm quarks $Q, \barQ$ and $D$ the
light quarks $q,\barq$.
In diagram Fig.\@ \ref{freebild} all spin and color states are degenerate
in the final state, and the sum over spin,
flavor and color quantum numbers gives a factor 1 for $I_S$ and $I_{FC}$.

As before, the spin-spin interaction is small and we consider the potential
contribution to ${\cal M}_{fi}$ only.
For ${\cal M}_{fi}^{\text{free}}$ we obtain
\be
M_{fi}^{\text{free}} (\bP,\bP',\tilde\bp_C,\tilde\bp_D)=&
 I_S^V& I_{FC}^V{\cal N}(s,\tilde\bp_C,\tilde\bp_D) H^V({\bf Q})
\nn
& \times& \left[\Phi^{*A}_0(\tilde\bp_C+{\textstyle{\frac{\bf Q}{2}}})-
                  \Phi^{*A}_0(\tilde\bp_C-{\textstyle{\frac{\bf Q}{2}}})\right]
 \nn
&\times & \left[ \Phi^{*B}_0(\tilde\bp_D +{\textstyle \frac{{\bf Q}}{2}})-
                \Phi^{*B}_0(\tilde\bp_D-{\textstyle{\frac{{\bf Q}}{2}}})
   \right],
\ee
with  $\Phi^A$ and $\Phi^B$ from Eq.\@ (\ref{1S}), $A=(Q\barQ),\, B=(q \barq)$
and ${\bf Q}=\bP-\bP'$. ${\cal N}(s,\tilde\bp_C,\tilde\bp_D)$ is given
according to Eq.\@ (\ref{calN}) by
\be
{\cal N}^2(s,\tilde\bp_C,\tilde\bp_D)&=&
     \frac{1}{s^2}\{s^2-(m_\psi^2-m_\pi^2)^2\}
\left\{s^2-(4(m_Q^2+\tilde\bp_C^2)-4(m_q^2+\tilde\bp_D^2))^2\right\}.
\ee
Inserting this into Eq.\@ (\ref{sigmaifs}) we obtain the
cross section into one definite momentum configuration
$\sigma^{\text{free}}_{fi}(s,\tilde\bp_C,\tilde\bp_D)$.
The total cross section $\sigma^{\text{free}}(s)$ of the process
$J/\psi+\pi \to Q+\barQ+q+\barq$ is given by integrating over all possible
relative momenta $\tilde \bp_C$ and $\bp_D$
\be
\sigma^{\text{free}}(s)&=&
   \int  \frac{d^3\tilde\bp_C}{(2 \pi)^3}
  \int \frac{d^3\tilde\bp_D}{(2 \pi)^3}
\sigma^{\text{free}}_{fi}(s,\tilde\bp_C,\tilde\bp_D).
\ee
The integration is restricted by energy conservation to
$0 \leq p_D^2 \leq \{\sqrt{s}/2-(m_Q^2+p_C^2)^{1/2}\}^2-m_q^2$ and
$0 \leq p_C^2 \leq (\sqrt{s}/2-m_q)^2-m_Q^2$.
Our result of the "perturbative" breakup cross section as a function of
$E_{rel}^{cms}$ is shown in Fig.\@ \ref{free2bild}. It does not exhibit a peak
close to threshold,
but starts smoothly and increases monotonically with energy. This is
analogous to what has been calculated in QCD perturbation theory
\cite{kharzeev}.

At low relative energies, the cross section of this process is small and it
does not contribute to the \j disintegration. The comparison of the two
cross sections, shown in Figs.\@ \ref{sigmabild} and \ref{free2bild},
demonstrates the importance of the correlation of the quarks in the final
state. It has the consequence of a strong enhancement close to threshold
where the relative momenta of the outgoing quarks are small and correlations
between them are most pronounced.

We emphasize at this point that the quark exchange reaction into correlated
final state mesons does not proceed via intermediate free quark states.
Therefore, the only energy barrier encountered in this process is the
reaction threshold, i.e. the mass difference between initial and final state
mesons. It is understood in our approach as the difference of the respective
binding energies, which is overcome by kinetic energy of the initial mesons.
However, we stress that no intermediate energy barrier is present
in this nonperturbative approach. This has to be seen in contrast
to a perturbative calculation,
where such a barrier occurs and where a nonperturbative mechanism, such as a
tunneling process, has to be invoked additionally.
\subsection{Inelastic cross sections in the strange sector}
We briefly look at the related processes involving strangeness instead of
charm. The meson--meson reaction in this case is
$ \phi + \pi \to K + \bar K $,
 which is, however, experimentally not
accessible. Instead, we look at the baryonic reactions in the strange sector
corresponding to the ones relevant for charmonium. These are
\be
\text{a)}\qquad  K^- + p &\to \Lambda + X,
  \nn
\text{b)}\qquad  K^+ + p &\to \Lambda + X,
\label{Kprocesses}
\ee
and a review of the data is given in \cite{kdata1,kdata2}. The cross
section for reaction a) exhibits a strong peak at threshold and a
subsequent decrease with increasing energy, while the cross section for
reaction b) increases monotonically from threshold. At energies far above
threshold both cross sections reach the same asymptotic value.
The data qualitatively show exactly the behavior we expect. Process
a) is dominated by a simple quark exchange process as we considered before,
for which we
calculated a strong peak at threshold, while reaction b) requires a
hard $s\bar{s}$ production process since $K^+$ contains an $\bar{s}$ quark,
while an $s$ is needed for the $\Lambda$. Therefore, reaction b) does not
show an enhancement at threshold. Although only being qualitatively, this
strongly supports the approach presented here.
%
\section{Dissociation kinetics in a pion gas}
\label{kinetics}
In this section, we consider the relaxation of the charmonium fluctuation by
string-flip processes in a dense hadronic medium such as the pion gas
produced in a high energy nucleus-nucleus  collision. To obtain the
suppression of the bound $Q\barQ$ states, we fold the energy dependent
absorption cross section calculated in the previous section with a thermal
pion distribution which is chosen in a way to describe the pion multiplicity
and shape of the rapidity dependence in the same reactions where the \j is
measured as well.

The time evolution of the \j distribution is described by the
Boltzmann equation \cite{blaizot,prorok2}
\be
    \frac{\partial}{\partial t}f_\psi ({\bf r},\bp_\psi,t)
    +\frac{\bp_\psi}{E_\psi} \nabla f_{\psi}({\bf r},\bp_\psi,t)
   &=&-f_{\psi} ({\bf r},\bp_\psi,t)
   \int \frac{d^3 \bp_{\pi}}{(2 \pi)^3}  f_{\pi}(\bp_{\pi},{\bf r},t)
\nn
  &&\times
         \sigma_{\text{abs}}[s(\bp_\psi,\bp_\pi)]
          j(\bp_\psi,\bp_\pi),
\label{boltzmann}
\ee
where $s(\bp_\pi,\bp_\psi)$ is the center of mass energy and
\be
\label{j(p)}
   j(\bp_\psi,\bp_\pi)&=&
  \frac{\sqrt{[E_\psi(p_\psi)E_\pi(p_\pi)-\bp_\psi \cdot
 \bp_\pi]^2-m_\psi^2m_\pi^2}}
   { E_\psi(p_\psi) E_\pi(p_\pi)}
\ee
is the flux of pions in the rest frame of the $J/\psi$ (see App.
\ref{appendixa}). Due to the small number of $Q \barQ$ pairs,
the inverse process of \j production in $D \bar D$ scattering is neglected,
and the influence of the considered reaction on the pion distribution is
negligible.
The solution of Eq.\@ (\ref{boltzmann}) for an initial \j distribution
$f_\psi({\bf r},\bp_\psi,t_0)$ reads
\be
   f_\psi ({\bf r},\bp_\psi,t)&=&f_\psi ({\bf r}-{\bf v}_\psi t,\bp_\psi,t_0)
       \exp \biggl[-\int_{t_0}^t d\, t'
         \int  \frac{d^3 \bp_\pi}{(2 \pi)^3}
\Bigr. \nn &&\Bigl.
    \times  f_\pi ({\bf r}-{\bf v}_\psi (t-t'),\bp_\pi,t')
         \sigma_{\text{abs}}[s(\bp_\psi,\bp_\pi)]j(\bp_\psi,\bp_\pi) \biggr].
\ee

We are interested in the time evolution of the total number of $J/\psi$'s
resulting from  the absorption by the breakup process considered in
section \ref{diss}.
For a qualitative estimate we consider the survival probability of a \j in
a uniform thermal pion gas. In this case the integration over the space
coordinate {\bf r} can be performed and the resulting
momentum distribution of meson $i$ is given by
\be
n_i (\bp_i,t) &=& \int d^3 {\bf r} f_i ({\bf r},\bp_i,t).
\ee
The Boltzmann equation
(\ref{boltzmann})
simplifies to
\be
  \label{ratengleichung}
    \frac{\partial n_{\psi}(\bp_\psi,t)}{\partial t}&=& -n_{\psi}(\bp_\psi,t)
       \frac{1}{\tau(\bp_\psi,t)},
\ee
where  the relaxation time $\tau(\bp_\psi,t)$
 is defined in the rest frame of the pion gas by
\be
 \tau(\bp_\psi,t)^{-1}&=& \left<\sigma_{\text{abs}} v_{rel}\right>_{n_\pi}
 \rho_{\pi}(t),
\ee
with the pion density
\be
\rho_\pi(t)&=&\int  \frac{d^3\bp_\pi}{(2\pi)^3} n_\pi(\bp_\pi,t).
\ee
 The brackets denote the average over the pion distribution $n_\pi$ which may,
 in general, be time dependent.
\be
\left<\sigma_{\text{abs}} v_{rel}\right>_{n_\pi (t)}&=&\frac{1}{\rho_\pi (t)}
        \int  \frac{d^3\bp_\pi}{(2 \pi)^3} n_\pi(\bp_\pi ,t)
  \nn
   &&   \times \sigma_{\text{abs}}[s(\bp_\psi,\bp_\pi)]
        j(\bp_\psi,\bp_\pi).
\ee

In order to give a quantitative estimate of the relaxation time for
the \j distribution in a dense pion gas, we consider the specific
example of a thermal pion distribution in equilibrium as given by the Bose
distribution
\be
\label{piverteilung}
      f_{\pi}(E_{\pi},T)
      &=&3 \left(\exp\left[(E_{\pi}-\mu)/{T}\right]-1\right)^{-1},
\ee
where the factor 3 stands for the pion multiplicity.
The temperature $T$ and chemical potential $\mu$ may be chosen as time
dependent for modeling the evolution of density and energy density
in nucleus-nucleus collisions. For the chemical potential of pions
we use the value $\mu_\pi= 126$ MeV with which the experimental heavy ion data
can be well reproduced \cite{mupi}.
The temperature range from 120 to 210 MeV corresponds to pion densities
from 0.22 to 0.84 $\text{fm}^{-3}$.

We show our result for the thermal averaged cross section
$\left<\sigma_{\text{abs}} v_{rel}\right>_T$ for different
temperatures and momenta of the \j relative to the pion gas center of mass
in Fig.\@ \ref{sigmavbild}. This quantity gives the mean capability of a pion
to dissolve a $J/\psi$ which is moving with momentum $\bp_{\psi}$ through the
pion gas. For low momenta of the $J/\psi$, it corresponds to a small cross
section since the relative energy exceeds the reaction threshold only in
few collisions. The cross section then rises with increasing momentum of the
$J/\psi$.
For all values of $\bp_\psi$, the absorption cross section increases with
increasing temperature of the pion gas.
The resulting mean life time of a \j moving through a pion gas, defined
in Eq.(36), is plotted in Fig.\@ \ref{taubild} as a function of the pion
temperature for different \j momenta.
In comparison to previous assumptions or phenomenological calculations
\cite{prakash,ftacnik89,gavin-satz}, we find a rather strong absorption of
the $J/\psi$. This is caused by the enhancement of the energy dependent cross
section
when quark-antiquark correlations in the final state are taken into account.

\section{Discussion of results in view of the experiments}
\label{experiment}
\subsection{\j absorption in a pion gas}
For a qualitative discussion of this result, we compare the calculated
relaxation time $\tau$ with the mean life time of a hadronic fireball which
is measured e.\@g.\@ by interferometry in the NA35 experiment and found to be
in the range of 5 fm/c for freeze-out temperatures $T\leq 150$MeV \cite{na35}.
This means that the relaxation time is of the same size as the lifetime of
the fireball. From our results, we conclude that \j dissociation in a dense
pion gas is capable of producing a rather large absorption. In particular, it
is large enough to describe the \j suppression as observed by NA38.

However, so far we have discussed an idealistic situation which is not met
in heavy ion collisions. First, the assumption of an equilibrium pion gas
without baryons and resonances is not realistic.
Second, in the early stage of the collision, the densities are
so high that a description by a free pion gas is not appropriate, and it is
uncertain at which time such a description becomes valid.
Third, as we discussed in detail in the previous section, the \j cannot be
regarded as a fully developed object from the very beginning. For the
latter two reasons, we overestimate the contribution of collisions with
pions to the \j absorption.
It is therefore interesting to analyze the variation of
the effective cross section
$\left< \sigma_{\text{abs}} v_{rel} \right>$ again as a function of the
radius of the $Q\barQ$ wave function  $\left< r^2 \right>_{Q\barQ}$.
For the free reaction $Q\bar{Q} +\pi \to D+ \bar{D}$, we had obtained a cross
section proportional to the radius squared of the $Q\bar{Q}$ pair. We find
that this relation still holds after averaging over the medium. The quantity
$\left< \sigma_{\text{abs}} v_{rel} \right>$ is shown in
Fig.\@ \ref{lccbild} for different temperatures of the surrounding pion gas
as a function of $\left< r^2 \right>_{Q\barQ}$.
It is approximately proportional to the mean squared $Q \barQ$ radius.
 Due to the symmetry of the quark exchange process we obtain the relation
\be
  \left< \sigma_{\text{abs}} v_{rel} \right> \propto
 \left< r^2 \right>_{Q\barQ}\left< r^2 \right>_{q\barq} \, ,
\ee
which can be regarded as a temperature-averaged version of
the Povh--H\"ufner relation \cite{povh}.

It has been emphasized on the basis of hadron--nucleus experiments
that there must be a considerable contribution to the absorption of \j on
nucleons \cite{gerschel2}.
Within our formalism, absorption of \j on nucleons such as the reaction
(22 c)
can be treated in an analog manner. We expect qualitatively a similar result
as we obtained for the breakup of $J/\psi$'s on pions.
With a choice of the time-dependent size of the $Q\bar{Q}$ system appropriate
for the respective kinematics, the hadronic absorption of charmonium on
nucleons should be well described by this approach, except for the very
high momentum region ($x_F \sim 1$) \cite{ger}.
In nucleus--nucleus collisions, we have
the additional absorption on the pion gas formed in the collision, on which
we concentrated in this work. With the proper kinematics, it is reduced from
the idealistic estimate above, but still to a value comparable to the
absorption on nucleons. Therefore, our results show that the two hadronic
absorption processes on pions and on nucleons, when taken together, are able
to account for the observed \j suppression.

\subsection{Heavy ion beams and inverse kinematics}
In the near future, experiments will take data with the lead beam at CERN.
It has been proposed to also study reactions in inverse kinematics, i.e.~a
heavy projectile on a light target. In this setup, the \j formed in the
hard collision will subsequently be taken over by the heavy projectile
nucleus at a low relative momentum. Thus, it is an ideal tool to study the
interaction of (almost) fully formed $J/\psi$'s on nucleons. From perturbative
calculations, no absorption is expected in this case due to the low density
of hard gluons in nuclear matter. Our prediction differs from this.
{}From our cross section, we expect a strong absorption of the \j as
compared to the case of pA collisions. The limitations of the model
we presented only allow us to make qualitative predictions.

We consider the situation of the NA38 experiments with a beam energy of 200 GeV
per nucleon, where dimuon pairs from
$J/\psi$'s can be detected in the rapidity window of
$2.8 \leq y_\psi \leq 4.0$. So we have in the mean $\bar y=3.4$.
We compare the case of proton beam on lead target (a: $p\, Pb \to \psi X$)
with a lead beam on a light target (b: $ Pb\, p \to \psi X$).
Both cases differ in the rapidity difference of the detected $J/\psi$'s and
the nucleons in the heavy ion. In the proton beam situation (a) we have a
rapidity difference of $ \Delta y=3.4$, which corresponds to a Lorentz factor
of $ \gamma_a=15$ and a
c.m.s.~energy of a nucleon and a $J/\psi$ of $\sqrt{s_a}=9.88\GeV $.
In the lead beam experiment with inverse kinematics (b),
we have $y_b=2.6,\, \gamma_b=6.8$ and  $\sqrt{s_b}=7.06 \GeV$.
The A dependence of the \j production cross section for a time dependent
absorption has been studied, among others, in Ref.\cite{blaizot3}.

We adopt a linear time evolution of the $Q \barQ$ radius before $\tau_{exp}$.
{}From our result, Eq.\@ (\ref{povhhuefner}), follows an evolution of
$\sigma_{\text{abs}}$ in terms of the $Q \barQ$ proper time $\tau$ as
\be
\sigma_{\text{abs}}(\tau,s)&=&
   \left\{ {\sigma_0(s) \frac{\tau^2}{\tau_{exp}^2}\, \quad \text{for } \tau
\leq \tau_{exp}}
 \atop
              {\sigma_0(s) \qquad \text{for } \tau > \tau_{exp}.}\right.
\ee

Let us introduce the ratio of the \j cross sections in direct (a) and inverse
(b) kinematical processes as
\be
R&=& \frac{\sigma_{pA\to\psi}}{\sigma_{Ap\to\psi}}.
\ee
Then, following Ref.\@ \cite{blaizot3}, we find for the quadratic time
dependence of the absorption cross section a ratio of
\be
 R & =&\exp\left[-\frac{A}{4\pi \tau_{exp}^2}\left(\frac{\sigma_0(s_a)}
       {\gamma_a^2 v_a^2}-\frac{\sigma_0(s_b)}{\gamma_b^2 v_b^2}
       \right) \right] \; ,
\ee
where $v_a \approx v_b \approx 1$ are the relative velocities of the \j
with respect to the scatterers.
We assume that initial state modifications and color octet
absorption \cite{mutzbauer} in both pA and Ap collisions occur at short
time scales inside the
nuclei such that these effects cancel when taking this ratio.
Inserting the values given above and using an energy independent cross
section of
$\sigma_0(s_a)=\sigma_0(s_b)=5$ mb, we obtain $R=1.22$. Concluding,
we expect a strong suppression of \j production in the inverse kinematical
regime as compared to the lead target.

An enhancement of the \j absorption cross section near threshold, as we
obtained for absorption on pions, leads to an even stronger suppression
in this case.
Taking, for example, $\sigma_0(s_b)=1.5 \sigma_0(s_a)$ gives $R=1.38$.
On the other hand, perturbative estimates suggest for this ratio a value of
$R=1$. Therefore, the experiment carried out in inverse kinematics is clearly
able to discriminate between the different models, and thus to indicate the
dominant physical processes.
%
\section{Conclusions}
\label{conclusion}
The aim of the present work was to establish a formalism in which the
absorption of \j mesons on hadronic matter is treated in a microscopic
approach. We describe mesons in a potential model, and consider
quark exchange reactions between two mesons as the model for inelastic
reactions such as open charm formation in a $J/\psi$--hadron collision.
For an exploratory calculation, we have used a Fermi-Breit
Hamiltonian and a Gaussian ansatz for the quark-antiquark wave functions.
In this model the energies of bound states are fairly well reproduced.
In addition, the model can still
be treated analytically and gives transparent results which we have
discussed in detail.

In this work, we concentrated on the reaction $J/\psi + \pi \to D +\bar{D}$
and calculated the cross section for the breakup reaction as a function
of the relative energy of the colliding mesons.
For comparison, we also considered the breakup reaction in free quark
states. The result of the latter calculation is similar to what is obtained in
the corresponding calculation in the framework of short distance QCD.
The comparison of both our results, as shown in Figures
\ref{sigmabild} and \ref{free2bild}, demonstrates
the importance of correlations of the quarks in the final state.

When considering correlated quarks, i.e. mesons, in the final state,
we find an enhanced cross section at the reaction threshold, that is, for low
relative momenta of the quarks in the final state. In general we are able to
describe a correlated $Q \barQ$ pair before it propagates to become  a
fully developed $J/\psi$.
For both kinds of final states, we find an increase of the absorption cross
section proportional to the square of the (time dependent) system size of
the $Q\bar{Q}$ state.

With a view to the application of this result to heavy ion collisions, we have
estimated the suppression of $J/\psi$'s in a dense pion gas in thermal
equilibrium.
Since this is an idealized situation, the results can only be considered
qualitatively. They suggest that in the
temperature region accessible to present experiments, non--perturbative
(string-type) correlations in the final state (asymptotically, these are
charmed hadrons) as described in the present approach are crucial for the
understanding of the \j suppression pattern. The enhanced cross section that
we obtain at the $D \bar{D}$ threshold is sufficient to explain the \j
suppression in the NA38 experiment as due to absorption by pions and nucleons.
This is in contrast to previous claims that \j suppression could only be
related to the quark-gluon plasma formation \cite{kharzeev}. We pointed out
that a measurement comparing $pA$ with $Ap$, in inverse kinematics, could
clearly distinguish between both physical pictures, since the one based on
perturbative calculations predicts basically
no suppression, while in the one presented here, a strong suppression is
expected.

The formalism that we have presented in this work is rather powerful.
A straightforward extension can simply accommodate higher states,
such as the charmonium states
($\chi_c,\psi'$), mesonic resonances ($\rho,\omega,\ldots$), and baryons
($N,\Delta$). We mentioned that similar
results are expected for charmonium absorption on nucleons.
Since quite a sizable contribution to the \j yield stems from $\chi$ decays,
also the $\chi - \pi$ cross section should be calculated.

An analysis of the $\psi'$ absorption cross section would be of particular
interest, since the ratio $\psi'$ to $J/\psi$ yields in $p$A and AB
collisions is supposed to be free of initial state effects. This ratio was
measured recently by NA38 \cite{quarkmatterna38}, and it seems to be a much
clearer probe of the state of matter than the \j to continuum signal.

As a further outlook, we mention also the possible application of the
present approach to the interaction of $J/\psi$'s with deconfined quark matter,
see e.g.\@ \cite{blaschkeroepke}. For this application the present
nonrelativistic calculation of the matrix element
${\cal M}_{fi}$ has to be improved by evaluations within a relativistic
potential model \cite{basti}, where the effects of chiral symmetry restoration
and quark deconfinement at finite temperature can be included.
\section*{Acknowledgment}
We thank J. H\"ufner, G. R\"opke and H.Satz for the permanent interest in this
work, S. Brodsky and E. Shuryak for engaged discussions and D. R\"ohrich for
information on the NA35 data. This work started in the Heidelberg-Rostock
collaboration and was supported by a grant from the Deutsche
Forschungsgemeinschaft (DFG) under contract  numbers Ro 903/7-1 and Hu 233/4-2.
%
\begin{appendix}
\section{Relativistic kinematics}
\label{appendixa}
We consider the two particle process $A+B \to C+D$.
The Mandelstam variables $s$, the center of mass energy of $A$ and $B$,
and $t$ are given by
\be
  s &=& ({\cal P}_A+{\cal P}_B)^2
    =(E_A+E_B)^2-(\bP_A+\bP_B)^2,
\label{mandels}
\\
t&=& ({\cal P}_A-{\cal P}_C)^2
    =(E_A-E_C)^2-(\bP_A-\bP_C)^2.
\ee
The flux $j$ of particle $A$ considered in the rest frame of $B$ is
\be
  j&=&v_{rel}\frac{{\cal P}_A {\cal P}_B}{E_A E_B}
\nn
   &=&\frac{ \sqrt{(s-s_+)(s-s_-)}}{2 E_{\psi} E_{\pi}}
\label{flux}
\\
\mbox{with}\qquad \qquad
  s_\pm &=& (m_A \pm m_B)^2.
\nonumber
\ee
$v_{rel}$ is the velocity of $A$ in the rest frame of $J/\psi$
\be
  v_{rel}^2  &=& 1-\frac{4(m_{A}m_{B})^2}{(s-m_{A}^2-m_{B}^2)^2} .
\label{vrel}
\ee
By inserting Eq.\@ (\ref{mandels}) into (\ref{flux}) one obtains
the form of Eq.\@
(\ref{j(p)}) for $j$, which depends on the three--momenta of both particles.
The following relations  between three--momenta (defined by Eq.(\ref{comdef}))
and Mandelstam
variables are valid in the center of mass frame of particles $A$ and $B$:
\be
 {P}^2(s)&=&
    \frac{1}{4s}\left\{ \left[s-(m_A^2+m_B^2)\right]^2-4m_A^2m_B^2\right\},
  \nn
{{P}'}^2(s)&=
 &\frac{1}{4s}\left\{\left[s-(m_C^2+m_D^2)\right]^2-4m_C^2m_D^2\right\},
 \nn
 2 \bP' \bP&=& 2P'P \cos \theta
  \nn
   &=&
         t-m_A^2-m_C^2+\frac{(s+m_A^2-m_B^2)(s+m_C^2-m_D^2)}{2s},
 \label{pcom}
\\
E_A(s) E_B(s) &=& \frac{1}{4s}\left(s^2-(m_A^2-m_B^2)^2\right),
\nn
E_C(s) E_D(s) &=& \frac{1}{4s}\left(s^2-(m_C^2-m_D^2)^2\right).
\ee
If four identical mesons with mass $m$ are considered, these formulae
simplify
\be
  {P}^2={{P}'}^2&=&\frac{1}{4}(s-4m^2),
 \nn
   2 \bP'\bP &=&t-2m^2+\frac{s}{2}.
  \label{pgleich}
\ee
The kinematical factor ${\cal N}(P,P')$ of Eq.\@ (\ref{Mfi}) is given
in the center of
mass frame by
\be
{\cal N}(s)&=& 4\sqrt{E_A E_B E_C E_D}
 \nn
      &=& \frac{1}{s} \sqrt{\left(s^2-(m_A^2-m_B^2)^2\right)
                                   \left(s^2-(m_C^2-m_D^2)^2\right)}.
   \label{calN}
\ee
\section{Fit of the meson spectrum}
\label{appendixb}
For an exact solution the mesonic wave functions have to be calculated by
solving the Schr\"odinger equation in the rest frame ($\bP_A=0$).
We use Gaussian wave functions of the form of Eq.\@ (\ref{1S}) as
approximating test functions. Then $\left|\Phi^A_0\right>$
depends only on one parameter $\lambda$, and is denoted by
 $\left|\Phi^A_\lambda\right>$. The best fit is found by using the
Ritz' variational principle. The Schr\"odinger equation reads
\be
        H \big| \Psi^A_0(\bp_a, \bp_{\bara }) \big>&=& m_A \big|
        \Psi^A_0(\bp_a,\bp_{\bara })
                \big>,
 \nn
       H &=& m_a +m_{\bara } +p_{a}^2/2m_a+p_{\bara }^2/2m_{\bara }
                   +H^I,
\ee
where $m_A$ is the mass of meson $A$ and $H^I=H^V+H^{SS}$ is the interaction
Hamiltonian given in Eq. \@(\ref{breitfermi}).
Only $H^{SS}$ gives a spin dependent contribution to the meson mass.
So we get for the mass of the $1S$ state
\be
   m_A&=&m_a +m_{\bara }+
          \frac{3(m_a+m_{\bara })}{16\lambda_{A} m_a m_{\bara }}
       -V_0\left(1+\frac{\lambda_{A}}{2x}\right)^{-3/2}
       -(\frac{3}{4}-S_A)\Delta M^{SS},
\ee
 where $S_A$ is the total spin of the meson $A$. Under the
assumption that the orbital component of the meson wave function is equal for
different spin states we can eliminate the spin dependent term.
Averaging over the spin and isospin states, the spin-spin
contribution to the Hamiltonian cancels and we get an averaged mass $m_A^{av}$
for $1S$ states $\left|\Psi(\lambda)\right>$ which depends on the
wave function parameter $\lambda$. For the ground state one has to fulfill
approximately the conditions
\begin{mathletters}
\label{Mav}
\be
      m_A^{av}(\lambda)&=&  \big<\Psi^A(\lambda)\big|H-H^{SS}\big|
             \Psi^A(\lambda)\big>
 \nn
              &=& \frac{3}{4}m_A^{S=1}(\lambda)+\frac{1}{4}m_A^{S=0}(\lambda)
 \nn
    &=&m_a +m_{\bara }+
          \frac{3(m_a+m_{\bara })}{16\lambda m_a m_{\bara }}
       -V_0\left(1+\frac{\lambda}{2x}\right)^{-3/2},
\ee
and
\be
     \left. \frac{\partial}{\partial \lambda_A}
              \big<\Psi^A(\lambda)\big|H-H^{SS}\big|\Psi^A(\lambda)\big>
     \right|_{\lambda=\lambda_A}
  &=&0
\ee
\end{mathletters}
for each of the quark pairings $q\-\barq$, $q\-\barQ, Q\-\barq$ and
$Q\-\barQ$
with appropriate parameters for quark masses $m_i$ and wave
function parameters $\lambda_{ij}$.
We determine the parameters in
 (\ref{ImpPot}) and (\ref{1S}) from the meson masses of
$\pi ,\varrho ,D, D^*,\eta_c$ and $J/\psi$ to \\
\be
\begin{minipage}{14em}
\be   m_Q&=& 1.84 \GeV,
 \nn
   m_q&=& 0.34 \GeV,
 \nn
   x&=& 1.47\GeV^{-2},
 \nn
   V&=& 1.24 \GeV.
\nonumber
\ee
\end{minipage}
\begin{minipage}{15em}
\be
   \lambda_{QQ}&=&0.755 \GeV^{-2},
 \nn
   \lambda_{qq}&=&3.05 \GeV^{-2},
 \nn
   \lambda_{Qq}&=&2.1 \GeV^{-2},
 \nn
 \text{ }_{\text{ }}
\nonumber
\ee
\end{minipage}
\label{parameter}
\ee
Table \ref{masstable} shows a comparison of the known meson
masses with the calculated masses with the model parameters (\ref{parameter}).
Since some masses
are uncertain, the averages are only approximately. The calculated root mean
squared radii of the states are given in the last column.
%
\section{Calculation of transition matrix elements}
\label{appendixc}
The Born matrix element ${\cal M}_{fi}$ has to be calculated by
integrating over all internal quark variables.
Here we demonstrate this for processes with exchange of antiquarks of
mesons A and B.
If identical quarks are involved, one has to add to each diagram the
corresponding ones decorated
with a fermion commutation operator (with negative sign, see
\cite{quarkexchange}).
The interaction can be written as sum of individual
interactions between the quarks $a, \bara , b$ and $\barb$, each of them
consisting of potential ($H^V$) and spin-spin ($H^{SS}$) contribution.
We only show the calculation of the diagram ${\cal M}^{V}_{a \barb}$ of
Fig.\@ \ref{feyn3}. Calculation of  the other diagrams  is analogous.
\subsection{Orbital factor of the matrix element}
First we consider the orbital factor of Eq.\@ (\ref{Va-b}),
\be
  I_{O,a \barb}^V(\bP,\bP')&=&
\quad \big<\Phi^A_\bP \Phi^B_{\text{-}\bP}\big|H^V_{a \barb}\big|
      \Phi^C_{\bP'} \Phi^D_{\text{-}\bP'} \big>.
\ee
In the center of mass frame the mesons A, B, C and D are moving with
three-momenta $\bP,-\bP,\bP'$ and $-\bP'$.
 From momentum conservation follows
$  \bp_a+\bp_{\barb}=\bp_c+\bp_{\barc},
  \bp_{\bara }=\bp_{\bard},
$ and $ \bp_{b}=\bp_{d}$.
Using these relations one can substitute six of the eight quark
variables and has to sum up over the two remaining $\bp_a$ and $\bp_c$.
\be
  I_{O,a \barb}^V(\bP,\bP')
 =\sum_{{\bp_a, \bp_{\bar a }, \atop \ldots}\atop \bp_d,\bp_{\bar d}}
  &&\big<\Phi^A_\bP | \bp_a\, \bp_{\bara } \big>
  \big<\Phi^B_{\text{-}\bP}|{\bf p}_b\, \bp_{\barb}\big>
      \big<\bp_a \,  \bp_{\barb}
      \big|H^V \big|\bc\, \bp_{\barc}\big>
   \nn
    &&\times \big<\bp_{\bara }\, \bp_b|\bp_{\bard}\, \bp_c \big>
      \big<\bp_c \, \bp_{\barc}|\Phi^C_{\bP'}\big>
      \big<{\bp_d}\, \bp_{\bard} |\Phi^D_{\text{-}\bP'}\big>
\nn
  = \ \sum_{{\bp}_a,{\bp_c}}^{ }&& \Phi^{*A}(\ba-(1\text{-}\eta_A) \bP)
         \Phi^{*B}(\ba-\bP'-\eta_{B} \bP)  H^V(\ba-\bc)
  \nn
         &&\times \Phi^C(\bc-(1\text{-}\eta_C) \bP')
         \Phi^D(\ba-\bP-\eta_{D} \bP').
\ee
The sums over $\bp_a$ and $\bp_c$ are replaced by integrals according to
 \be
  \sum_{\left| {\bf p}_i \right>} &\rightarrow&
   \int \frac{d^3 {\bf p}_i}{(2 \pi)^3}.
\label{sum-int}
\ee
In our model we use Gaussian wave functions and a Gaussian shape of the
potential.
In this case an  analytical expression for the orbital overlap matrix element
in the $(1S)+ 1S\to (1S)+(1S)$
process can be obtained,
\be
  I^V_{O\, a \barb}(P,P',z)&=&\big<\Phi^A_\bP \Phi^B_{\text{-}\bP}\big|
   H^V_{a \barb}\big|\Phi^C_{\bP'} \Phi^D_{\text{-}\bP'} \big>
\nn
&=&-64 V_0\left(\frac{8 x}{\pi}\right)^{3/2}
  (\lambda_A \lambda_B \lambda_C \lambda_D)^{3/4}
  \int d^3\ba \int d^3 \bc
  \nonumber \\
   && \times \exp \left[-2 \left\{
  \lambda_A(\ba-(1-\eta_A) \bP)^2+\lambda_B(\ba-\bP'-\eta_B \bP)^2+
  x(\ba-\bc)^2 \right\}\right]
 \nn
   &&\times \exp\left[-2\left\{\lambda_C(\bc-(1-\eta_C) \bP')^2
                       +\lambda_D(\ba-\bP-\eta_D \bP')^2
        \right\}\right]
\label{ortsant2}
\\
 &=& -K_{a \barb}
    \exp{\left[-\left(\alpha_1 {\rm P}^2+\alpha_2 {{\rm P}'}^2
                +\alpha_3 P P'z\right)\right]}.
\label{ortsantappendixabarb}
\ee
For the general case we give only the constant $\alpha_1$,
\be
 \alpha_1&=&2\left\{
       \lambda_A \lambda_B(\eta_A+\eta_B-1)^2+\lambda_A\lambda_C'(1-\eta_A)^2+
      \lambda_A\lambda_D(\eta_A)^2
 \right.
 \nn
   &&\left.
     +\lambda_B\lambda_C'\eta_B^2
       +\lambda_B\lambda_D(1-\eta_B)^2+\lambda_C'\lambda_D\right\}
     (\lambda_A+\lambda_B+\lambda'_C+\lambda_D)^{-1},
\ee
with $1/{\lambda'}_C=1/\lambda_C+1/x$.
The other constants have similar forms.

In the special case of the quark rearrangement reaction
(\ref{abc}c) in section \ref{diss} we have due to flavor conservation
\be
   m_a&=&m_{\bara } =m_c=m_{\bard}=m_{Q},
 \nn
  m_b&=&m_{\barb} =m_{\barc} =m_d=m_q.
 \ee
 We introduce the notation $\eta=\frac{m_Q}{m_Q+m_q}$.
The outgoing D mesons have the same radii,
\be
 \lambda_C &=&\lambda_D=\lambda_{Qq}.
\ee

Then the parameters of the capture diagram $I^{V}_{O,a\barb}$ simplify to
\be
 \alpha_1&=&\frac{2}{\Lambda_\alpha}\left\{
\frac{1}{4}(\lambda_{QQ}+\lambda_{qq})\left(\lambda_{Qq}'
            +\lambda_{Qq}\right)
            +\lambda_{Qq}\lambda_{Qq}' \right\},
     \nn
  \alpha_2&=&\frac{2}{\Lambda_\alpha}\left\{
     \left(\lambda_{Qq}+\lambda_{Qq}'\right)
             \left(\lambda_{QQ}\eta^2+\lambda_{qq} (1-\eta)^2\right)
             +\lambda_{QQ}\lambda_{qq} \right\},
      \nn
  \alpha_3&=&   \frac{2}{\Lambda_\alpha}\left(\lambda_{QQ}\eta-\lambda_{qq}
             (1-\eta)\right)
              \left(\lambda_{Qq} -\lambda_{Qq}'\right),
\label{alphas}
\\
  \Lambda_\alpha&=&\lambda_{QQ}+\lambda_{qq}+\lambda'_{Qq}+\lambda_{Qq},
\\
K_{a \barb}&=&
        V_0 \left(\frac{32 \pi  \lambda'_{Qq}}{\Lambda_\alpha}\right)^{3/2}
         (\lambda_{QQ} \lambda_{qq})^{3/4}=K_{\bara b}.
\label{Kabarb}
\ee
In a analogous way we obtain $I^V_{O\, ab}$,
\be
  I^V_{O\, a \barb}(P,P',z)&=&\big<\Phi^A_\bP \Phi^B_{\text{-}\bP}\big|
   H^V_{a b}\big|\Phi^C_{\bP'} \Phi^D_{\text{-}\bP'} \big>
\nn
 &=& -K_{ab}
    \exp{\left[-\left(\beta_1 {\rm P}^2+\beta_2 {{\rm P}'}^2
                +\beta_3 P P'z\right)\right]},
\label{ortsantappendixab}
\ee
where in the considered process
\be
\beta_1&=&\frac{2}{ \Lambda_\beta}\bigl[\frac{1}{4}\lambda_{QQ}
    (\lambda_{qq}+\lambda_{Qq}+2x)+\frac{1}{4}\lambda_{qq}(\lambda_{QQ}+
    \lambda_{Qq}+2x)
     +\lambda_{Qq}x\bigr],
\nn
\beta_2&=&\frac{2}{\Lambda_\beta}\left[\eta^2\lambda_{QQ}
     ( \lambda_{qq}+\lambda_{Qq}+2x)+(1-\eta)^2\lambda_{qq}
  (\lambda_{QQ}+\lambda_{Qq}+2x)+\lambda_{QQ}\lambda_{qq}x/\lambda_{Qq}\right],
\nn
\beta_3&=&\frac{2}{\Lambda_\beta}\left[\eta\lambda_{QQ}
         (\lambda_{qq}+\lambda_{Qq})+(1-\eta)\lambda_{qq}
          (\lambda_{QQ}+\lambda_{Qq})\right],
  \label{betas}
\\
\label{lambdabeta}
\Lambda_\beta&=&\left[(\lambda_{QQ}+
\lambda_{Qq}+x)(\lambda_{qq}+\lambda_{Qq}+x)-x^2 \right]/\lambda_{Qq},
\\
K_{ab}&=&V_0 \left(\frac{32 \pi x}{\Lambda_\beta}\right)^{3/2}
         (\lambda_{QQ} \lambda_{qq})^{3/4}=K_{\bara \barb}.
\label{Kab}
\ee
$P'$ is determined by $P$ from energy conservation.
The matrix element can be rewritten in kinematical variables $s$ and $t$,
see App.\@ \ref{appendixa}.
The orbital factors for the remaining diagrams $I^V_{O\, \bara b}$ and
$I^V_{O\, \bara \barb}$ are obtained by replacing $z$ by $-z$
in (\ref{ortsantappendixabarb}) and (\ref{ortsantappendixab}) because the
orbital wave functions of mesons $C$ and $D$ are identical.
%
\subsection{Spin factor}
\label{spinappend}
The spin wave functions for spin singlets
and spin triplets are given by Clebsch--Gordan coefficients
\be
  \bigl|\chi_S^{S_A,S_A^z}\bigr>&=&\sum_{{\bf s}_a,{\bf s}_{\bar a }}
       \left<s_a,s_a^z,
                      s_{\bara },s_{\bara }^z|S_A,S_A^z\right>
                      \left|s_a,s_a^z,s_{\bara },s_{\bara }^z\right>.
\\
I^V_{S}  &=&\big<\chi^A_S \chi^B_S\big|\chi^C_S \chi^D_S \big>
 \nn
  &=&\sum_{a,\bar a ,b \atop \ldots,d,\bar {d}}
      \chi^S_A ({\bf s}_a, {\bf s}_\bara ) \chi_B^S({\bf s}_b {\bf s}_\barb)
      \delta^{_{(S)}}_{a,c} \delta^{_{(S)}}_{\bara,\bard}
      \delta^{_{(S)}}_{b,d} \delta^{_{(S)}}_{\barb, \barc}\,
  {\bf 1}\, \chi^S_C({\bf s}_c {\bf s}_\barc)\chi^S_D({\bf s}_d {\bf s}_\bard).
\ee
The different values of this factor are summarized in Table \ref{ispintable}.
The spins of initial mesons and the sum of them are in the head of the table
and of the reaction products in the first column. The spin factor in the
potential interaction term of the
transition matrix element between these states can be read from this table.
For the $I^{SS}_S$ term one has to multiply these factors by the numbers
given in the right two columns for the diagrams ${\cal M}_{a \barb}$
and ${\cal M}_{\bara b}$.
The factors for ${\cal M}_{a b}$ and ${\cal M}_{\bara \barb}$ diagrams are
obtained in a similar way, see \cite{barnes}.
\subsection{Flavor-color factor}
The color singlet wave function for mesons is
\be
\chi_{C}^A  &= & \frac{1}{\sqrt{3}}
        \sum_{c_a, c_{\bar a}=1}^3 \delta^{_{(C)}}_{a,\bara}
\ee
and the flavor and color give only a combinatoric factor in the
Born matrix element.
{}From the color component we get an overall factor of 1/3 for
${\cal M}_{a \barb}$ and  ${\cal M}_{\bara b}$ and -1/3 for
${\cal M}_{a b}$ and ${\cal M}_{\bara \barb}$. This accounts for the fact that
only a third of all quark pairs, those of identical color, are able to
produce a color singlet in the quark exchange process.
This result differs by a factor 4/3  from Ref.\@ \cite{barnes}. This factor is
included in the effective form of our model interaction.
The light quark flavors $u, d$ are assumed to be degenerate,
\be I_{FC}   &=&\big<\chi^A_{FC} \chi^B_{FC}\big|\chi^C_{FC} \chi^D_{FC} \big>
 \nn
&=&\sum_{a,\bar a ,b \atop \ldots,d,\bar {d}}
      \chi_{FC}^A ( a, \bara ) \chi^B_{FC}(b\, \barb)
      \delta^{_{(F,C)}}_{{a},{c}} \delta^{_{(F,C)}}_{\bara , \bard}
      \delta^{_{(F,C)}}_{{b},{d}} \delta^{_{(F,C)}}_{\barb, \barc}
      \chi_{FC}^A(c\, \barc)\chi_{FC}^C(d\, \bard).
\ee
{}From the Kronecker deltas follow that only the quark line diagrams shown
in Fig.\@ \ref{feyn3} give nonzero contributions to the matrix elements.
All other possible diagrams are forbidden in our case of four different
quarks.
$I^{FC}$ gives $\pm 1/3$ if flavor conservation is fulfilled and zero if not.
\end{appendix}




 \begin{table}[p] 
\centerline{
\begin{minipage}[b]{40em}
 \caption{\label{ispintable}}
         Spin factors $I^{V}_{S}$ of the transition matrix elements
         for different spin states of the initial $(A,B)$ and
         final $(C,D)$ mesons.
         For the $I^{SS}_{S}$ factors one has to multiply each line by the
factors
         in the respective column $I^{SS}_{S\, a \barb}$ or
         $I^{SS}_{S\, \bara b}$.
\end{minipage}
          }
 \end{table}

\begin{table}[p]
\centerline{
\begin{minipage}[b]{40em}
          \caption{
 \label{channels}}
           The spin factors for different final channels $f=(C,D)$
            in the quark exchange process $J/\psi+\pi \to C+D$.
            They follow directly
            from Table \ref{ispintable}.
            $\sigma_0, s_0$ and $a$ are parameters for
            the fit formula Eq.\@ (\ref{fit}).
\end{minipage}
}
\end{table}

\begin{table}[p]
\centerline{
\begin{minipage}[b]{40em}
          \caption{ \label{masstable}}
            Meson mass spectrum according to the formula
            Eq.\@ (\ref{Mav}) with the fitted model parameters
           (\ref{parameter}) in comparison with the spin averaged
           experimental masses. In the last column, the root mean
           squared radii of the mesonic state are given.
\end{minipage}
}
\end{table}


\begin{figure}[tb]
\centerline{
     \begin{minipage}[b]{40em}
     \caption[The quark exchange diagrams contributing to ${\cal M}_{fi}$
           in first Born approximation]
          {          \label{feyn3}}
           Contributions to the quark exchange matrix element
            ${\cal M}_{fi}$ in the Born approximation:
             ${\cal M}_{a \barb},
            {\cal M}_{\bara  b}$ (capture) and ${\cal M}_{a b}$,
           ${\cal M}_{\bara \barb}$ (transfer).
           Each interaction line represents the sum of
            potential and spin-spin interaction.
      \end{minipage}
}
\end{figure}

\begin{figure}[htb]
\centerline{
     \begin{minipage}[b]{40em}
     \caption{ \label{sigmabild}}
          Cross section for different channels of inelastic rearrangement
          reactions of $J/\psi$ and $\pi$ into $D$ mesons.
      \end{minipage}
             }
\end{figure}

\begin{figure}[htb]
\centerline{
     \begin{minipage}[b]{40em}
     \caption{          \label{freebild}}
          Diagram for the disintegration reaction of \j and $\pi$ into four
           free quarks.
      \end{minipage}
             }
\end{figure}

\begin{figure}[htb]
\centerline{
     \begin{minipage}[b]{40em}
          {
     \caption{\label{free2bild}}
          Cross section for the reaction $J/\psi+\pi \to Q+\barQ+q+\barq$.
          This result is comparable to the perturbative calculation
          of Kharzeev and Satz \protect{\cite{kharzeev}}.
          }
      \end{minipage}
             }
\end{figure}

\begin{figure}[htb]
\centerline{
     \begin{minipage}[b]{40em}
          {
     \caption{\label{sigmavbild}}
           Thermal averaged cross section $\left<\sigma v_{rel}\right>_T$
          for $J/\psi+\pi \to D(1S)+\bar D(1S)$
          for a $J/\psi$ moving with momentum $|\bp_{\psi}|$ through a gas of
          pions with a chemical potential $\mu=126$ MeV
          at different temperatures $T$.
        }
      \end{minipage}
             }
\end{figure}

\begin{figure}[htb]
\centerline{
     \begin{minipage}[b]{40em}
          \caption{ \label{taubild}}
          Mean life time $\tau$ of a \j in a pion gas as a function of
        the temperature $T$ for different momenta $|\bp_\psi|$ of the \j with
         respect to the pion gas center of  mass.
      \end{minipage}
  }
\end{figure}

\begin{figure}[htb]
\centerline{
     \begin{minipage}[b]{40em}
          {
     \caption{\label{lccbild}}
           Dependence of the thermal averaged cross section
          $\left<\sigma v_{rel}\right>_T$
          on the mean squared radius of the $Q \barQ$ wave function
         for a $Q \barQ$ pair at rest in an equilibrium pion gas with
         chemical potential $\mu=126$ MeV at different temperatures $T$.
         The nearly linear dependence for small $\left<r^2_{QQ}\right>$
         corresponds to the Povh--H\"ufner relation \protect{\cite{povh}}.
        }
      \end{minipage}
             }
\end{figure}


\begin{thebibliography}{99}

   \bibitem{qm93}
{\it Quark Matter 93}, Proceedings of the Tenth Conference on
Ultrarelativistic Nucleus-Nucleus Collisions, Borl\"ange, Sweden
1993, Nucl. Phys. {\bf A 533}, 1c,  (1994).


   \bibitem{matsui}
T. Matsui and H. Satz,
       Phys. Lett. {\bf B 178}, 416 (1986).

  \bibitem{na38data}
M.C. Abreu et al. (NA38 collab.),
   Z. Phys {\bf C 38}, 117 (1988).

  \bibitem{karsch}
 F. Karsch and H. Satz, Z. Phys. {\bf C 51}, 209 (1991).

   \bibitem{huefner}
J. H\"ufner, Y. Kurihara and H.J. Pirner,
     Phys. Lett. {\bf B 215}, 218 (1988).

    \bibitem{gavin2}
S. Gavin and M. Gyulassy,
   Phys Lett. {\bf B 214}, 241 (1988).

   \bibitem{na3}
J. Badier et al. (NA3), Z. Phys. {\bf C 20}, 101 (1983).

   \bibitem{alde}
D. M. Alde et al.,
         Phys. Rev. Lett. {\bf 66}, 133 (1991).

   \bibitem{gerschel1}
C. Gerschel and J. H\"ufner,
      Z. Phys {\bf C 56}, 171 (1992).

     \bibitem{kharzeev}
D. Kharzeev and H. Satz, Phys. Lett. {\bf B 334}, 155 (1994).


    \bibitem{wittmann}
R. Wittmann and U. Heinz,
   Z. Phys. {\bf C 59}, 77 (1993).

   \bibitem{lattice1}
C. Bernard, T.A. DeGrand, C. DeTar, S. Gottlieb, A. Krasnitz, M.C. Ogilvie,
R.L. Sugar and D. Toussaint,
   Phys. Rev. Lett. {\bf 68}, 2125 (1992).

   \bibitem{lattice2}
K.D. Born, S. Gupta, A. Irb\"ack, F. Karsch, E. Laermann,
B. Peterson and H. Satz,
     Phys. Rev. Lett. {\bf 67}, 302 (1991).

     \bibitem{koch-brown}
V. Koch, E.V. Shuryak, G.E. Brown and A.D. Jackson,
    Phys. Rev. {\bf D 46}, 3169 (1992).

   \bibitem{blaschkeroepke}
D. Blaschke, G. R\"opke and H. Schulz,
    Phys. Lett. {\bf B 233}, 434 (1989).

   \bibitem{brod}
J.F. Gunion, S.J. Brodsky and R. Blankenbecler, Phys. Rev.
    {\bf D8}, 287 (1973)
and {\bf D12}, 3469 (1975).

   \bibitem{qex1}
J. Weinstein and N. Isgur, Phys. Rev. Lett. {\bf 48}, 659 (1982);
        Phys. Rev. {\bf D 27}, 588 (1983).

   \bibitem{qex2}
F. Lenz, J.T. Londergan, E.J. Moniz, R.Rosenfelder, M.Stingl and  K. Yazaki
 Ann. Phys. (N.Y.) 170, 65 (1986).

   \bibitem{qex3}
A.M. Green and G.Q. Liu, Nucl. Phys. {\bf A 500}, 529 (1989).

   \bibitem{qex4}
F. Fernandez, A. Valarce, U. Straub and A. Faessler,
          J. Phys. {\bf G 19}, 2013 (1993).

   \bibitem{barnes}
T. Barnes and E. S. Swanson,
          Phys. Rev {\bf D 46}, 131 (1992).

   \bibitem{quarkexchange}
D. Blaschke and G. R\"opke,
     Phys. Lett. {\bf B 299}, 332 (1993).

   \bibitem{lucha}
W. Lucha, F.F. Sch\"oberl and D. Gromes,
          Phys. Rep. {\bf 200}, 127 (1991).


\bibitem{barnes2}
T. Barnes and E.S. Swanson,
   Phys. Rev. {\bf C 49}, 1166 (1994)

   \bibitem{vogtsystematics}
R. Vogt, S. J. Brodsky and P. Hoyer,
      Nucl. Phys. {\bf B 360} 67 (1991).


   \bibitem{prorok3}
D. Prorok and L. Turko,
       Z. Phys. {\bf C 61}, 109 (1994).

   \bibitem{frankfurt}
L. Frankfurt and M. Strikman, Prog. Part. Nucl. Phys. {\bf 27}, 135 (1991).

   \bibitem{blahuef}
D. Blaschke and J. H\"ufner, Phys. Lett. {\bf B 281}, 364 (1992).

   \bibitem{quack}
E. Quack,
    Nucl. Phys. {\bf B 364}, 321 (1991).

   \bibitem{povh}
B. Povh and J. H\"ufner,
       Phys. Lett. {\bf B 245}, 653 (1990).

   \bibitem{kdata1}
J. Whitmore, Phys. Rep. {\bf 27}, 187 (1976).

   \bibitem{kdata2}
P. Wright et al., Nucl. Phys. {\bf B 189}, 421 (1981).

   \bibitem{blaizot}
J.-P. Blaizot and J.-Y. Ollitrault,
    Phys. Rev. {\bf D 39}, 232 (1989).

   \bibitem{prorok2}
D. Prorok,
       Phys. Lett. {\bf B 275}, 465 (1992).

   \bibitem{mupi}
K. Kataja and P.V. Ruuskanen,
   Phys. Lett. {\bf B 243}, 181 (1990).

   \bibitem{prakash}
R. Vogt, M. Prakash, P. Koch and T. H. Hansson,
            Phys. Lett. {\bf B 207}, 263 (1988).

   \bibitem{ftacnik89}
J. Ft\'a\v cnik, P. Lichard, J. Pi\v sutov\'a and J. Pi\v s\'ut,
            Z. Phys. {\bf C 42}, 139 (1989).

   \bibitem{gavin-satz}
S. Gavin, H. Satz, R.L. Thews and R. Vogt,
     Z. Phys {\bf C 61}, 351 (1994).

   \bibitem{na35}
D. R\"ohrich et al. (NA35 collab.),
   Nucl. Phys. {\bf A 566}, 35c (1994).

   \bibitem{gerschel2}
C. Gerschel and J. H\"ufner,
       Phys. Lett. {\bf B 207}, 253 (1988).

   \bibitem{ger}
For a recent overview of the high $x_F$ physics of charmonium, see for
instance C. Gerschel, J. H\"ufner and E.Quack, ``Phenomenological analysis
of the $x$--distribution for \j production on nuclei'',
preprint HD--TVP--94--1, Heidelberg 1994

   \bibitem{blaizot3}
J.-P. Blaizot and J.-Y. Ollitrault,
     Phys. Lett. {\bf B 217}, 386 (1989).

   \bibitem{mutzbauer}
G. Piller, J. Mutzbauer, W. Weise,
      Nucl. Phys. {\bf A 560} (1993) 437.


   \bibitem{quarkmatterna38}
C. Louren\c{c}o, in \cite{qm93}

    \bibitem{basti}
S. Schmidt, D. Blaschke and Y. Kalinovsky,
     Phys. Rev. {\bf C 50}, 435 (1994).


\end{thebibliography}
\end{document}